\documentclass[12pt]{article}
\usepackage{amsmath,amssymb}

\DeclareMathOperator{\tr}{tr}

\newcommand{\CC}{\mathbb{C}}
\newcommand{\Rr}{\mathbb{R}}
\newcommand{\Ss}{\mathbb{S}}
\newcommand{\LL}{\mathcal{L}}

\begin{document}
\begin{center}
    {\Large An alternative interpretation of the Weinberg-Salam model.}
\end{center}
\begin{center}
    L.~D.~Faddeev\\
        St.~Petersburg Dept. of Steklov Mathematical
        Institute,\\
        Russian Academy of Sciences
\end{center}
\begin{flushright}
    \emph{Si nous ne trouvons pas des choses agr\'eables,\\
    nous trouverons du moins des choses nouvelles.}\\
    Cacambo to Candide before finding Eldorado
\end{flushright}

    My wife and me found this phrase while reading on the Crimean beach during
    free time Voltaire's ironic description of the Candide's adventures. (What
    can you do on the beach but collecting pebbles and reading.) We could not
    help noting, that this sentence reflects feelings of the large part of
    physical community in wake of the results on LHC. In my talk, based on the
    joint paper with A.~Niemi and M.~Chernodub
\cite{FNC},
    I shall present a nonorthodox approach to WS model, proposing a new
    interpretation for the Higgs field. In particular I shall argue, that
    masses of vector bosons could be suplemented without use of the Higgs
    potential.

    The paper
\cite{FNC}
    was produced by email correspondence and the version of A.~N. and M.~Ch.
    was published. Here I shall use my approach, which is fully equivalent
    to
\cite{FNC}
    and add some personal remarks.

    The new interpretation concerns only bosonic part of WS model, so I shall
    consider only lagrangian for complex scalar dublet
$ \Phi=(\phi_{1},\phi_{2}) $,
    abelian vector field
$ Y_{\mu} $
    and SU(2) Yang-Mills triplet
$ B^{a}_{\mu}, a=1,2,3 $
\begin{equation*}
    \LL= \left(\nabla_{\mu}\Phi,\nabla_{\mu}\Phi\right) + \frac{1}{4g^{2}}
	B^{a}_{\mu\nu} B^{a}_{\mu\nu} +\frac{1}{4g'^{2}} Y_{\mu\nu}^{2} ,
\end{equation*}
    where
\begin{align*}
    \nabla_{\mu}\Phi =& \partial_{\mu} \Phi +\frac{i}{2} Y_{\mu} \Phi
	+ B^{a}_{\mu} t^{a} \Phi \\
    B^{a}_{\mu\nu} =& \partial_{\mu} B^{a}_{\nu} - \partial_{\nu} B_{\mu}
	+ \epsilon_{abc} B^{b}_{\mu} B^{c}_{\nu} \\
    Y_{\mu\nu} =& \partial_{\mu} Y_{\nu} - \partial_{\nu} Y_{\mu}
\end{align*}
    with
$ t^{a} = \frac{i}{2} \tau^{a}$,
$ \tau^{a} $ --- Pauli matrices,
$ (\cdot,\cdot) $ --- hermitian scalar product in $ \CC^{2}$, 
$ g $ and $ g' $ --- coupling constants.

    I do not introduce selfinteraction for the scalar field
$ \Phi $.
    The interpretation below is exactly based on this omission. In fact all my
    friends among theoretical physicists hate
$ \phi^{4} $
    interaction, it is not asymptotically free and quite possible dissapears
    under proper renormalization.

    The lagrangian
$ \LL $
    has U(2) gauge invariance with parameters: matrix
$ \Omega $
    from SU(2) and real function
$ \omega $
\begin{align*}
    \Phi^{\Omega} = & \Omega\Phi, \quad \Phi^{\omega} = e^{i\omega} \Phi,\\
    B^{\Omega}_{\mu} =& \Omega B_{\mu}\Omega^{-1} - 
	\partial_{\mu} \Omega \Omega^{-1}, \quad B^{\omega}_{\mu} = B_{\mu} ,\\
    Y_{\mu}^{\Omega} =& Y_{\mu} , \quad  Y_{\mu}^{\omega} = Y_{\mu} -
	2\partial_{\mu} \omega .
\end{align*}
    Here
$ B_{\mu} = B_{\mu}^{a}t^{a} $.
    The idea of
\cite{FNC}	
    is to make the change of variables, which leads to the gauge invariant
    degrees of freedom. Before presenting the explicite formulas I shall give
    a geometric reason for them. The ``target'' for the field
$ \Phi $ is
$ \CC^{2} $ or
$ \Rr^{4} $
    if we count real components. In radial coordinates
$ \Rr^{4} $
    can be presented as
$ \Rr_{+} \times \Ss^{3} $
    and furthermore
$ \Ss^{3} $
    is (almost) the same as SU(2). The SU(2) degrees of freedom, realized as
    matrix
$ g $,
    allow to introduce gauge transformation of the Yang-Mills field, leaving
    the gauge invariant vector field. In
\cite{FNC}
    one of realization of this idea was proposed.
    I am sure, that this comment is not original, a similar considerations
    were discussed for example long ago in
\cite{Sh}.
    However, I believe that the interpretation given below is new.

    Let us proceed. We should extract the SU(2) degrees of freedom from 
$ \Phi $
    in most convenient way. Here is my proposal.
    First write
$ \Phi $
    as
\begin{equation*}
    \Phi = \rho \chi ,
\end{equation*}
    where
$ \rho $
    is a positive function and
$ \chi $ --- normalised as follows
\begin{equation*}
    (\chi,\chi) = \bar{\chi}_{1} \chi_{1} + \bar{\chi}_{2} \chi_{2} = 1.
\end{equation*}
    The the matrix
\begin{equation*}
    g=
\begin{pmatrix}
    \chi_{1} & - \bar{\chi}_{2} \\
    \chi_{2} & \bar{\chi}_{1}
\end{pmatrix}
    =
\begin{vmatrix}
\chi, \sigma \bar{\chi}
\end{vmatrix}
\end{equation*}
    is unimodular and unitary. Here
\begin{equation*}
    \sigma = \frac{1}{i} \tau_{2} =
\begin{pmatrix}
    0 & -1 \\
    1 & 0
\end{pmatrix} .
\end{equation*}
    Doublet
$ \sigma \bar{\chi} $
    transforms under the nonabelian gauge transformation exactly as
$ \chi $, indeed
\begin{equation*}
    \bigl(\sigma \bar{\chi}\bigr)^{\Omega} = \sigma \bar{\Omega} \bar{\chi}
	= -\sigma \bar{\Omega} \sigma \sigma \bar{\chi}
	= \Omega \sigma \bar{\chi} ,
\end{equation*}
    where I use the properties
$ \sigma^{2} = -I$ and
$ \sigma \bar{\tau} \sigma = \tau $
    for all Pauli matrices. Thus the whole matrix
$ g $
    transforms as
\begin{equation*}
    g^{\Omega} = \Omega g .
\end{equation*}
    The abelian transformation is different for
$ \chi $ and
$ \bar{\chi} $,
    so that
\begin{equation*}
    g^{\omega} = g
\begin{pmatrix}
    e^{i\omega} & 0 \\
    0 & e^{-i\omega}
\end{pmatrix}
    = g e^{i\omega \tau_{3}} .
\end{equation*}
    We see, that covariant derivative of
$ g $
    assumes the form
\begin{equation*}
    \nabla_{\mu} g = \partial_{\mu} g + \frac{i}{2} Y_{\mu} g \tau_{3}
	+ B_{\mu} g
\end{equation*}
    and
$ |\nabla_{\mu}\Phi|^{2} $
    can be rewritten as
\begin{equation*}
    |\nabla_{\mu}\Phi|^{2} = \frac{\rho^{2}}{2}
	\tr \bigl((\nabla_{\mu} g)^{*} (\nabla_{\mu} g) \bigr) 
	+ \partial_{\mu} \rho \partial_{\mu} \rho .
\end{equation*}
    Introduce the new vector field
\begin{equation*}
    W_{\mu} = g^{*} B_{\mu} g + g^{*} \partial_{\mu} g .
\end{equation*}
    It is easy to check, that
\begin{align*}
    W_{\mu}^{\Omega} = & W_{\mu} ,\\
    W_{\mu}^{\omega} = & e^{-i\omega\tau_{3}} W_{\mu} e^{i\omega\tau_{3}}
	+ i \partial_{\mu}\omega \tau_{3}
\end{align*}
    and direct calculation shows, that
\begin{equation*}
    \frac{1}{2} \tr \bigl((\nabla_{\mu} g)^{*} (\nabla_{\mu} g) \bigr)
	= \frac{1}{4} \bigl(Y_{\mu}^{2} + 2Y_{\mu}W_{\mu}^{3}
	 +(W_{\mu}^{a})^{2}\bigr)
    = \frac{1}{4} \bigl(Z_{\mu}^{2} + W_{\mu}^{+} W_{\mu}^{-}\bigr) ,
\end{equation*}
    where
\begin{equation*}
    Z_{\mu} = Y_{\mu} + W_{\mu}^{3}
\end{equation*}
    and we introduce the components of field
$ W_{\mu} $
\begin{align*}
    W_{\mu} =& \frac{i}{2} (W_{\mu}^{1}\tau_{1} + W_{\mu}^{2}\tau_{2}
        + W_{\mu}^{3}\tau_{3}) \\
    W_{\mu}^{\pm} =& W_{\mu}^{1} \pm W_{\mu}^{2} .
\end{align*}
    In these components
\begin{equation*}
    (W_{\mu}^{\pm})^{\omega} = e^{\pm 2i\omega} W_{\mu}^{\pm}, \quad
    (W_{\mu}^{3})^{\omega} = W_{\mu}^{3} + 2\partial_{\mu} \omega ,
\end{equation*}
    so that abelian vector field
$ Z_{\mu} $
    is completely gauge invariant and
$ W_{\mu}^{\pm} $ and
$ W_{\mu}^{3} $
    behave under abelian transformation as charged and abelian gauge vector
    field, correspondingly.

    The vector part of the lagrangian can be rewritten as
\begin{equation*}
    \frac{1}{4g'^{2}} Y_{\mu\nu}^{2} + \frac{1}{4g^{2}} (
	W_{\mu\nu}^{3} + H_{\mu\nu})^{2} + \frac{1}{4g^{2}}
	(\nabla_{\mu}W_{\nu}^{+} -\nabla_{\nu}W_{\mu}^{+})
	(\nabla_{\mu}W_{\nu}^{-} -\nabla_{\nu}W_{\mu}^{-}),
\end{equation*}
    where
\begin{align*}
    W_{\mu\nu}^{3} =& \partial_{\mu} W_{\nu}^{3} - \partial_{\nu} W_{\mu}^{3}\\
    H_{\mu\nu} =& \frac{1}{2i} (W_{\mu}^{+}W_{\nu}^{-} -W_{\mu}^{-}W_{\nu}^{-})
\end{align*}
    and
\begin{equation*}
    \nabla_{\mu} W_{\nu}^{\pm} =\partial_{\mu}W_{\nu}^{\pm} \pm 
	i W_{\mu}^{3} W_{\nu}^{\pm} .
\end{equation*}
    To express the sum of quadratic forms of
$ Y_{\mu\nu} $ and
$ W_{\mu\nu}^{3} $ via
$ Z_{\mu} $
    it is convenient to introduce the combination
\begin{equation*}
    A_{\mu} = \frac{1}{g^{2}+g'^{2}}(g'^{2}W_{\mu}-g^{2}Y_{\mu})
\end{equation*}
    such that
\begin{equation*}
    \frac{1}{4g'^{2}} Y_{\mu\nu}^{2} + \frac{1}{4g^{2}} (W_{\mu\nu}^{3})^{2} =
	\frac{1}{4(g^{2}+g'^{2})} Z_{\mu\nu}^{2} +
	\frac{g^{2}+g'^{2}}{4g^{2}g'^{2}} A_{\mu\nu}^{2}
\end{equation*}
    with as usual
\begin{equation*}
    Z_{\mu\nu} = \partial_{\mu}Z_{\nu} -\partial_{\nu}Z_{\mu} , \quad
    A_{\mu\nu} = \partial_{\mu}A_{\nu} -\partial_{\nu}A_{\mu} .
\end{equation*}
    The abelian vector field
$ A_{\mu} $
    transforms in the same way as
$ W_{\mu}^{3} $
\begin{equation*}
    A_{\mu}^{\omega} = A_{\mu} -2\partial_{\mu} \omega
\end{equation*}
    and enters into lagrangian via
$ \frac{1}{4e^{2}} A_{\mu\nu}^{2} $ with
$ e^{2} = \frac{g^{2}g'^{2}}{g^{2}+g'^{2}}$.
    It is clear, that
$ A_{\mu} $
    can be interpreted as electromagnetic field with electric charge given by
$ e $
    the 
$ \omega $-action
    having meaning of electromagnetic gauge transformation.

    Thus in new variables the list of fields consists of real positive field
$ \rho $,
    neutral abelian fields
$ Z_{\mu} $ and
$ A_{\mu} $
    and charged vector field
$ W_{\mu}^{\pm} $.
    The lagrangian assumes the form
\begin{align*}
    \LL =& \partial_{\mu} \rho \partial_{\mu} \rho +\frac{\rho^{2}}{4}
	(Z_{\mu}^{2}+W_{\mu}^{+}W_{\mu}^{-}) 
    + \frac{1}{4g^{2}} (\nabla_{\mu}W_{\nu}^{+}-\nabla_{\nu}W_{\mu}^{+})
	(\nabla_{\mu}W_{\nu}^{-}-\nabla_{\nu}W_{\mu}^{-}) \\
    +& \frac{1}{4(g^{2}+g'^{2})} Z_{\mu\nu}^{2} +
	\frac{1}{4e^{2}} A_{\mu\nu}^{2} 
    + \frac{2}{4g^{2}}(H_{\mu\nu},W_{\mu\nu}^{3})
	+\frac{1}{4g^{2}}H_{\mu\nu}^{2} ,
\end{align*}
    where 
$ W_{\mu}^{3} $
    should be changed to its expression via
$ Z_{\mu} $ and
$ A_{\mu} $.
    Now we can begin discussion.

    The change of variables, which I undertook, eliminates 3 degrees of
    freedom, leaving positive field
    $ \rho $
    instead of four real components of scalar
    $ \Phi $.
    The functional measure
\begin{equation*}
    d\mu = \prod_{x} d\phi_{1} \, d\bar{\phi}_{1} \, d\phi_{2} \,
	d\bar{\phi}_{2} \, dY_{\mu} \, dB_{\mu}^{a}
\end{equation*}
    used before gauge fixing, looks in new variables as
\begin{equation*}
    d\mu = \prod_{x} \rho^{2} d\rho^{2} \, dZ_{\mu} dW_{\mu}^{+} \, dW_{\mu}^{-}
	\, dA \, dg ,
\end{equation*}
    where
$ \prod_{x} dg $
    is volume of the gauge group, which is completely separated from measure
    without any gauge fixing. We are just to drop it to write nonsingular
    functional integral.

    We see, that
$ \rho^{2} $
    is not an ordinary scalar field. Besides being positive, it enters the
    functional integral with local factor. This requires some interpretation.
    In particular the reason for nontrivial expectation value
\begin{equation*}
    <\rho^{2}> = \Lambda^{2} ,
\end{equation*}
    supplying mass to vector fields
$ Z_{\mu} $ and
$ W_{\mu}^{\pm} $,
    must be elucidated.

    In
\cite{FNC}
    we proposed to interpret
$ \rho^{2} $
    as conformal factor of the metric in space-time
\begin{equation*}
    g_{\mu\nu} = \rho^{2} \delta_{\mu\nu} .
\end{equation*}
    Indeed, in 4-dimensional space-time we have
$ \sqrt{g} = \rho^{4} $,
    and contravariant vector and tensors have factors
\begin{equation*}
    \chi^{\mu} =\rho^{-2} \chi_{\mu} , \quad \chi^{\mu\nu} 
	=\rho^{-4}\chi_{\mu\nu},
\end{equation*}
    so that
$ \rho^{2}Z_{\mu}^{2} = Z_{\mu}Z^{\mu}\sqrt{g} $,
$ H_{\mu\nu}^{2} =H_{\mu\nu}H^{\mu\nu} \sqrt{g}$, etc.
    Moreover the scalar curvature is given by
\begin{equation*}
    R = \frac{1}{6} \frac{\partial_{\mu}\rho \partial_{\mu}\rho}{\rho^{4}}
	+ \text{divergence} .
\end{equation*}
    Finally, the Christoffel's symbols, entering the definition of the field
    strengths
$ A_{\mu\nu} $,
$ Z_{\mu\nu} $,
$ W_{\mu\nu}^{\pm} $
    via covariant derivatives, cancel due to antisymmetry.
    Thus the lagrangian
    can be rewritten in manifestly covariant form.

    In this interpretation it is natural to require, that
\begin{equation*}
    \rho^{2} |_{r\to\infty} \rightarrow \Lambda^{2}
\end{equation*}
    at spacial infinity to maintain the asymptotical flatness. Parameter
$ \Lambda^{2} $
    enters as a new parameter of the model.

    An alternative argument for
    nontrivial expectation value
$ \Lambda^{2} $
    looks as follows. The field
$ \rho $
    enters the lagrangian either via derivatives or being multiplied by
    another field.
    So the classical vacuum configuration, given by
\begin{equation*}
    \rho^{2} = \Lambda^{2} , \quad Z_{\mu}=0, \quad W_{\mu}^{\pm} = 0, \quad
	A_{\mu} = 0
\end{equation*}
    is degenerate. The choice of particular value of
$ \Lambda^{2} $
    corresponds to the concrete choice of the vacuum. All this looks as
    noncompact analogy of ferromagnetism.

    Thus, one way or another we see, that the nonzero expectation value for
    the
$ \rho^{2} $
    can be envoked without the Higgs potential. The fundamental question which
    remains, is the origine of the excitations for the field
$ \rho $.
    In both interpretations the most natural answer is massless scalar ---
    analogy of dilaton in the first interpretation or kind of Goldstone mode in
    the second.

    I hope, that more experienced phenomenologist can consider seriously this
    hypothesis.


\begin{thebibliography}{0}
\bibitem{FNC}
    M.~N.~Chernodub, L.~Faddeev, A.~Niemi, Non-abelian Supercurrents and
    Electroweak Theory,
    UUITP-04-08, ITEP-LAT-2008-10, arXiv:0804.1544 [hep-th].
\bibitem{Sh}
    V.~V.~Vlasov, V.~A.~Matveev, A.~N.~Tavkhelidze, S.~Y.~Khlebnikov and
    M.~E.~Shaposhnikov, Fiz. Elem. Chast. Atom Yadra {\bf 18} (1987) 5.
\end{thebibliography}
\end{document}